\PassOptionsToPackage{dvipsnames}{xcolor}
\documentclass[longbibliography,nofootinbib,aps,prx,superscriptaddress,twocolumn]{revtex4-2}

\usepackage{amsmath,amsfonts,amssymb,amsthm,dcolumn,bm,qcircuit,appendix,mathtools,thmtools,thm-restate,algorithm,algpseudocode,mathrsfs,pgfplots,soul,lipsum,graphicx,braket,enumitem,caption,subcaption,ragged2e}

\usepackage{hyperref}

\usepackage[
  labelfont=bf,
  labelsep=period,
  justification=justified,
  format=plain,
  singlelinecheck=false
]{caption}

\usepackage{pgfplots}
\usepgfplotslibrary{patchplots}
\usetikzlibrary{3d, decorations.pathmorphing}
\usepackage{tikz,tkz-euclide,tikz-3dplot}
\pgfplotsset{compat=newest}
\usepgfplotslibrary{colormaps}
\usepgfplotslibrary{patchplots}

\usetikzlibrary{matrix,fit,backgrounds,3d,arrows, decorations.pathreplacing,shapes.geometric,3d, calc}

\def\BibTeX{{\rm B\kern-.05em{\sc i\kern-.025em b}\kern-.08em
    T\kern-.1667em\lower.7ex\hbox{E}\kern-.125emX}}

\newtheorem{theorem}{Theorem}
\newtheorem{lemma}{Lemma}

\newtheorem{remark}{Remark}
\newtheorem{definition}{Definition}

\numberwithin{proposition}{section}
\numberwithin{theorem}{section}
\numberwithin{lemma}{section}
\numberwithin{corollary}{section}
\numberwithin{remark}{section}
\numberwithin{definition}{section}
\numberwithin{equation}{section}

\DeclareMathOperator{\Tr}{Tr}
\newcommand{\xbf}{\textbf{x}}

\newcommand{\Ibb}{\mathbb{I}}

\newcommand{\Rbb}{\mathbb{R}}

\newcommand{\norm}[1]{\left\lVert#1\right\rVert}

\usepackage{xcolor}
\usepackage{hyperref}

\hypersetup{
    colorlinks=true,
    linkcolor=MidnightBlue,
    filecolor=magenta,
    urlcolor=teal,
    pdfpagemode=FullScreen,
    citecolor=OliveGreen
}

\begin{document}
\title{Quantum Algorithm for Estimating Ollivier-Ricci Curvature }
\author{Nhat A. Nghiem}
\email{{nhatanh.nghiemvu@stonybrook.edu}}
\affiliation{Department of Physics and Astronomy, State University of New York at Stony Brook, \\ Stony Brook, NY 11794-3800, USA}
\affiliation{C. N. Yang Institute for Theoretical Physics, State University of New York at Stony Brook, \\ Stony Brook, NY 11794-3840, USA}
\author{Linh Nguyen}
\email{{linh.nguyen@famu.edu}}
\affiliation{Department of Mathematics, Florida A\&M University, Tallahassee, FL
32301, USA}

\author{Tuan K. Do}
\email{{ktdo@ucsb.edu}}
\affiliation{Department of Mathematics,  University of California, Santa Barbara, CA
93106, USA}

\author{Tzu-Chieh Wei}
\email{{tzu-chieh.wei@stonybrook.edu}}
\affiliation{Department of Physics and Astronomy, State University of New York at Stony Brook, \\ Stony Brook, NY 11794-3800, USA}
\affiliation{C. N. Yang Institute for Theoretical Physics, State University of New York at Stony Brook, \\ Stony Brook, NY 11794-3840, USA}

\author{Trung V. Phan}
\email{{tphan@natsci.claremont.edu}}
\affiliation{Department of Natural Sciences, Scripps and Pitzer Colleges, \\ Claremont Colleges Consortium, Claremont, CA 91711, USA}

\begin{abstract}
We introduce a quantum algorithm for computing the Ollivier–Ricci curvature, a discrete analogue of the Ricci curvature defined via optimal transport on graphs and general metric spaces. This curvature has seen applications ranging from signaling fragility in financial networks to serving as basic quantities in combinatorial quantum gravity. For inputs given as a point cloud with pairwise distances, we show that our algorithm can achieve an exponential speed-up over the best-known classical methods for two particular classes of problem. Our work is another step toward quantum algorithms for geometrical problems that are capable of delivering practical value while also informing fundamental theory.
\end{abstract}

\maketitle

\section{Introduction}

\ \ 

The \textit{manifold hypothesis} conjectures, with substantial empirical support, that any high-dimensional dataset can be represented by a much lower-dimensional manifold~\cite{fefferman2016testing,gorban2018blessing,brown2022verifying}. By stripping away redundant coordinates -- often artifacts of noise or constraints~\cite{sternad2018s} -- one can recover the true local degrees of freedom, i.e., the intrinsic dimensionality~\cite{amsaleg2015estimating}. By constructing local geometry -- typically via a nearest-neighbor graph~\cite{eppstein1997nearest} and pairwise distances -- we can infer the global topology of the underlying manifold and extract local geometric information such as curvature~\cite{lee2006riemannian}, whose bottlenecks or singularities flag anomalies~\cite{grover2025curvgad}. This approach for big data analysis has many practical applications, and perhaps best known through quantitative finance, where the curvature on asset-correlation graphs are found to be a good indicator for market fragility~\cite{sandhu2016ricci,samal2021network} and potential crash magnitude \cite{boginski2006mining,chi2010network}. Among discrete estimators of curvatures, the \textit{Ollivier–Ricci curvature} (ORC) \cite{ollivier2007ricci,ollivier2009ricci,SanchezGarciaGherghe2024ORCFragility} has often emerged to be the most useful: it is defined via optimal transport between neighboring probability measures on the graph \cite{santambrogio2015optimal}, associating curvature directly with how information moves across the network. For the same reason, ORC fits naturally with combinatorial quantum gravity \cite{trugenberger2017combinatorial,van2021ollivier,kelly2022convergence,trugenberger2023combinatorial}, where distances and Markovian dynamics create emergent geometry. As ORC gains traction, developing efficient methods to compute it becomes essential, including novel approaches that go beyond classical algorithms.

Quantum computation has seen significant progress since its early proposals~\cite{feynman2018simulating, benioff1980computer}. By leveraging the intrinsic nature of quantum physics, including entanglement and superposition, a quantum computer can store and process information indistinguishably from a classical one. As such, it holds great promise for solving problems beyond the classical limit. Early attempts~\cite{deutsch1985quantum, deutsch1992rapid, shor1999polynomial, grover1996fast} have, to some extent, revealed such promise by showing that quantum computers can compute the property of black-box functions, search the database, and factorize integers, which can be done with considerably fewer resources than the best-known classical algorithms \cite{bauer2020quantum, cerezo2021variational}. Over time, the notion of quantum computing has reached almost every corner of scientific computing. Notable examples include solving linear systems~\cite{harrow2009quantum, childs2017quantum, wossnig2018quantum}, Hamiltonian simulation~\cite{berry2007efficient,berry2012black,berry2014high,berry2015hamiltonian,berry2015simulating, lloyd1996universal,childs2018toward, gerritsma2010quantum, babbush2018low}, solving differential equations~\cite{berry2014high, arrazola2019quantum, childs2021high, liu2021efficient, krovi2023improved}, machine learning $\&$ artificial intelligence~\cite{schuld2018supervised, schuld2019evaluating, schuld2019machine, schuld2019quantum,schuld2020circuit, lloyd2013quantum, havlivcek2019supervised}, etc. 

Recently, quantum algorithmic techniques have been applied in the context of topological data analysis (TDA), igniting a series of efforts~\cite{lloyd2016quantum, berry2024analyzing, schmidhuber2022complexity, nghiem2023quantum,nghiem2025new, hayakawa2022quantum, gyurik2024quantum,ubaru2021quantum}. In a parallel effort, the work~\cite{nghiem2025quantum} explores the potential of quantum computing for geometrical data analysis (GDA). While built on different mathematical grounds, TDA and GDA offer powerful frameworks for analyzing large-scale data, capable of revealing intrinsic properties of data in realistic scenarios where noise and sampling error often intervene. As outlined in the aforementioned works, quantum computers can offer an efficient solution to many problems in the TDA and GDA context, thus suggesting a broader potential for application, going beyond what was shown in the context of scientific computing, simulation, machine learning, etc. 

In this work, following these developments -- particularly, the attempt in~\cite{nghiem2025quantum} -- we extend the reach of quantum computing to a closely relevant problem in the context of GDA. Specifically, we consider the problem of estimating ORC, given the input as a point cloud with pairwise connectivities and pairwise distances. This input model is the same as~\cite{nghiem2025quantum}; however, whereas~\cite{nghiem2025quantum} estimates local scalar curvature via a diffusion-based approach, in this work, we adopt an optimal-transport-based estimator.

Our work is organized as follows. In Section~\ref{sec: classicalalgorithm}, we provide a formal description of the key objective and related assumptions. In the same section, we describe the classical solution to the key objective and state our main result, followed up by a comparison to show the advantage of our quantum method. Section~\ref{sec: quantumalgorithm} contains our main proposal for the quantum algorithm for estimating the ORC, which splits into two subsections \ref{sec: case1} and \ref{sec: case2} to handle two cases, corresponding to two different types of input conditions. Finally, conclusion is given in Section \ref{sec: conclusion}.

\section{Classical Algorithm for Calculating the Ollivier-Ricci Curvature}
\label{sec: classicalalgorithm}
Consider a graph $G = (V,E)$, where $V$ is the set of vertices and $E$ is the set of edges. Each edge is represented by the pair of vertices it connects, $(a,b) \in E$ with $a,b \in V$, 
and is assigned a distance $d(a,b)$. The edges are undirected, so that 
$(a,b) \equiv (b,a)$. The \textit{neighbors} of a vertex $a \in V$ are defined as the set of all vertices connected to $a$ by an edge, i.e. $\mathcal{N}_a = \{b \in V | (a,b) \in E \}$. For any pair of vertices $a',b' \in V$, the \textit{graph, or geodesic distance} between them, denoted $d_G(a',b')$, is defined as the length of the shortest path in $G$ connecting $a'$ and $b'$. Explicitly,
\begin{equation}
d_G(a',b') = \min_{\substack{(v_1,v_2),(v_2,v_3),\ldots\\ (v_{k-2},v_{k-1}),(v_{k-1},v_k) \in E \\ v_1=a',\, v_k=b'}} \ 
\sum_{i=1}^{k} d(v_i,v_{i+1}) \ ,
\end{equation}
where the minimum is taken over all possible paths in the graph $G$ with endpoints $a'$ and $b'$. 

\begin{figure}[tb!]
    \centering
    \includegraphics[width=0.9\linewidth]{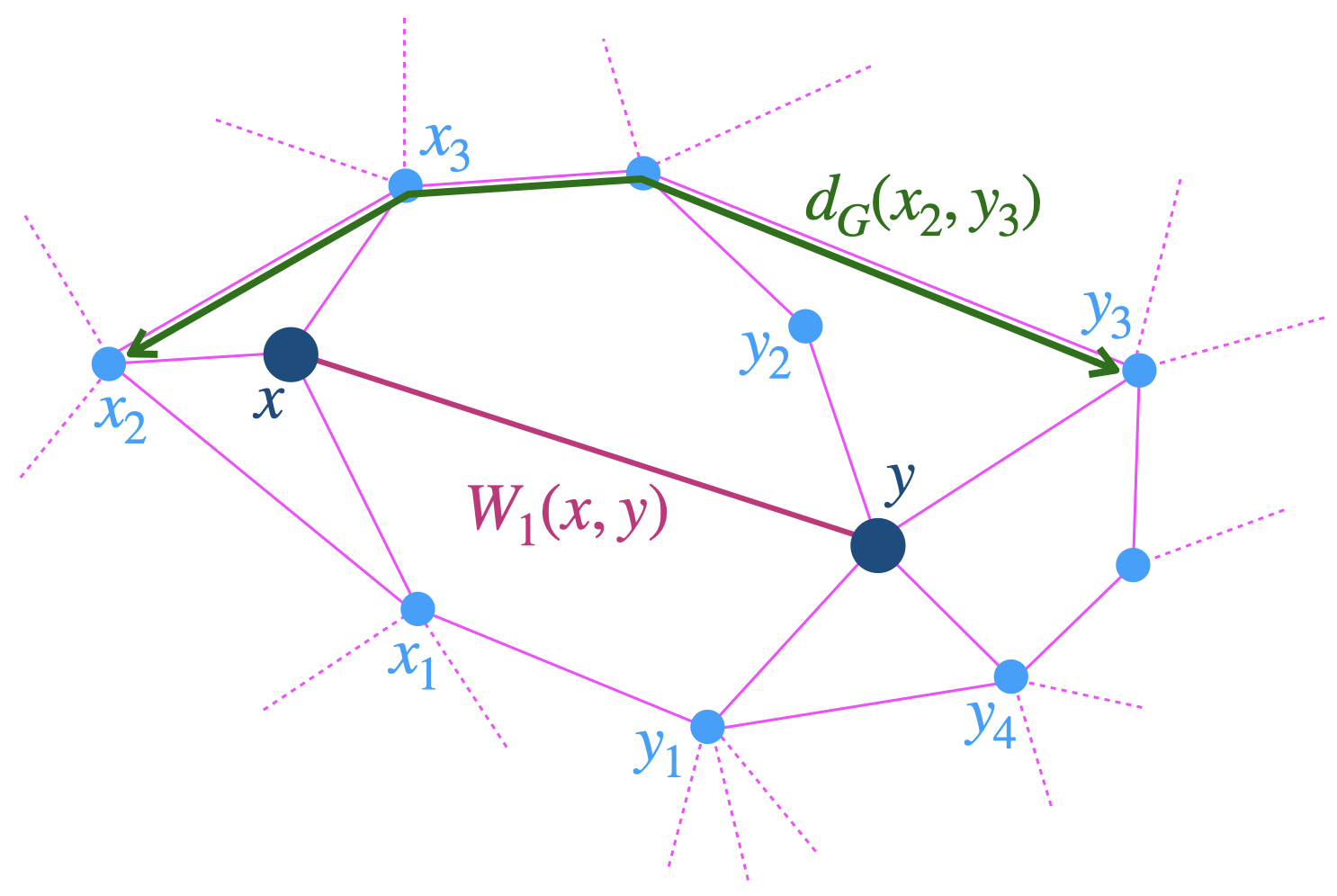}
    \caption{\textbf{Illustration of the setup used to compute the Earth Mover distance associated with an edge in a graph.} The graph's vertices are in blue, and its edges are in magenta. Here we consider the edge $(x,y)$, in which the neighbors of $x$ excluding $y$ are  $\{ x_1, x_2, x_3\}$ ($p=3$) and the neighbors of $y$ excluding $x$ are $\{ y_1, y_2, y_3, y_4\}$ ($q=4$). We show the graph distance $d_G(x_2,y_3)$ between points $x_2$ and $y_3$ in green. By optimizing the expression given in Eq.~\eqref{optimal_transport}, we arrive at the Earth Mover distance $W_1(x,y)$, which then allows us to calculate the ORC $\gamma(x,y)$ associated with the edge $(x,y)$ using Eq. \eqref{orc}.}
    \label{fig:orc}
\end{figure}

An ORC value is associated with each edge $(x,y) \in E$, denoted by $\gamma(x,y)$, and can be calculated with the Earth Mover (1-Wasserstein) distance $W_1(x,y)$ as follows \cite{saidi2024recovering}:
\begin{equation} 
\gamma(x,y) = 1 - \frac{W_1(x,y)}{d_G(x,y)} \ .
\label{orc}
\end{equation}
To calculate $W_1(x,y)$,  we need to focus on the local configuration near the edge $(x,y)$, as illustrated in Fig.~\ref{fig:orc}. Assume that the set of $x$-neighbors \textit{excluding} $y$ has $p$ points and the set of $y$-neighbors \textit{excluding} $x$ has $q$ points, i.e. $\mathcal{N}_x \setminus \{ y\} = \{ x_1, x_2, ..., x_p \}$ and $\mathcal{N}_y \setminus \{ x \} = \{ y_1, y_2, ..., y_q \}$. Then the value of $W_1(x,y)$ is the minimum cost of reconfiguring mass in vertices in $\mathcal{N}_x \setminus \{ y\}$, which we further assume to sum up to 1 and be uniformly distributed, to another uniform distribution of mass in vertices in $\mathcal{N}_y \setminus \{ x\}$. Concretely, denote by $\gamma_{ij}$ (where $i=1,2,...,p$ and $j=1,2,...,q$) the \textit{transport variable} that represents the amount of mass transferred from $x_i$ to $y_j$, which incurs a cost of $d_G(x_i, y_j)\gamma_{ij}$. We solve the following linear program with $p \times q$ variables and $p + q$ constraints:
\begin{align}
\label{optimal_transport}
    W_1(x,y) = \min_{\{\gamma_{ij}\}}  &\sum^p_{i=1}\sum^q_{j=1} d_G(x_i,y_j) \gamma_{ij}, \\
   \forall j:   \sum_{i=1}^p \gamma_{ij} &= 1/q,\\
     \forall i:  \sum_{j=1}^q \gamma_{ij} &= 1/p,\\ 
   \  \forall i,j: \gamma_{ij} &\geq 0. 
\end{align}
As a demonstration, in Appendix \ref{app:ORC_calc}, we calculate $\gamma(x,y)$ in the graph illustrated in Fig. \ref{fig:orc}, assuming that all edges are unit-length. We note that the definition of $W_1(x,y)$ varies, e.g., in \cite{saucan2019discrete,math8091416}, the summation above does include $y$ as neighborhood of $x$ (and vice versa). However, either definition is practically fine and does not lose the generality of our subsequent construction. 

In general, a linear program does not admit a closed form solution and needs to be solved algorithmically.  An open question in the field of optimization that has resisted progress for several decades is whether linear programming can be solved in strongly polynomial time. All known algorithms are either worst-case exponential (in the number of variables, for fixed dimension linear programming is strongly polynomial) or algebraic/numerical and thus depend on the bit complexity of the input~\cite{chvatal1983linear, dantzig2002linear}.

The particular category that our linear program computes $W_1(x,y)$ falls into is the structured \textit{optimal transport} problem, for which there exist strongly polynomial algorithms.  We give a brief survey of some of these and refer the reader to the references listed for more details.
\begin{enumerate}
    \item Orlin's strongly polynomial time combinatorial exact algorithm that runs in $O(pq(p + q)\log(p+q))$ for finding the min-cost flow on an uncapacitated complete bipartite graph $K_{p, q}$~\cite{orlin1988faster}.
    \item A recent breakthrough gives an $O\left((pq)^{1 + o(1)}\right)$ algorithm for optimal transport with polynomially bounded integral demands and supplies (into which we can easily transform our linear program, by scaling both demands and supplies by a factor of $pq$)~\cite{chen2025maximum}.
    \item When $p = q$, the 1-Wasserstein optimal transport problem becomes a linear assignment (matching) problem. To see this, rescale the variables by letting $\Gamma_{ij} = p\,\gamma_{ij}$. Then $\sum_{j=1}^p \Gamma_{ij} = 1$ and $\sum_{i=1}^p \Gamma_{ij} = 1$, so the matrix $\Gamma \in \mathbb{R}^{p\times p}$ is a square \textit{doubly stochastic matrix}~\cite{marshall1979inequalities}. The objective~\ref{optimal_transport} becomes $\frac{1}{p}\sum_{i=1}^p\sum_{j=1}^p d_G(x_i,y_j)\,\Gamma_{ij}$ which is equivalent (up to a constant factor, in the context of a linear program) to $\sum_{i,j} d_G(x_i,y_j)\Gamma_{ij}$.

By the classic Birkhoff theorem~\cite{birkhoff1946three}, every doubly stochastic matrix can be expressed as a convex combination of permutation matrices (matrices with exactly one 1 in each row and column). Equivalently, the permutation matrices are precisely the vertices of the Birkhoff polytope, which contains all doubly stochastic matrices. We are to optimize a linear objective function over the set of all doubly stochastic matrices and since in any linear program, if an optimal solution exists and the optimum value is finite (which is the case for our problem since the Birkhoff polytope is compact), then at least one optimal solution lies at a vertex of the feasible polytope~\cite{chvatal1983linear, dantzig2002linear}. Thus it suffices to find an optimal permutation matrix. In this case, an optimal transport plan sends all the mass at each vertex to exactly one other vertex, so the problem reduces to a balanced matching problem, solvable by the Hungarian algorithm. (However, the optimal value of the objective function $W_1(x,y)$ still depends on the cost structure and does not admit a closed-form expression~\cite{kuhn1955hungarian}.)

    We note that while there are other approaches with better asymptotic running times, the Hungarian algorithm has a small constant in the big $O$ and works well in practice with moderately sized instances, along with plentiful off-the-shelf implementations that are highly optimized, so the Hungarian algorithm remains the standard in many applications. 
    \item When the distances between the supply and demand nodes are \textit{decomposable}, i.e., $d_{ij} = e_i + f_j$ or $d_{ij} = e_i + f_j + C$ ($C$ is a constant), the solution is remarkably simple, as any feasible transport is optimal~\cite{santambrogio2015optimal}. For instance, when $G$ is a tree, since there is only one path from any $x_i$ to any $y_j$: $(x_i,x), (x, y), (y, y_j)$, we always have $d_G(x_i, y_j) = d_G(x_i, x) + d_G(x, y) + d_G(y, y_j)$, and therefore we have 
    \begin{align}
        W_1(x, y) =  \frac{1}{p}\sum_{i=1}^pd_G(x_i, x) + d_G(x, y) + \frac{1}{q}\sum_{j=1}^qd_G(y, y_j).
    \end{align}
   The classical algorithm in this case just needs to perform the summation, which takes $\mathcal{O}(p+q)$ time. 
    \end{enumerate}

As can be seen above, the complexity of the classical algorithms, and hence their complexities vary case by case, as linear programming in general is difficult and a closed-form solution does not exist. In this work, we pay particular attention to (1) when $G$ forms a tree and (2) when $p=q$. In these cases, there are certain patterns in the solutions that we could exploit to construct the corresponding quantum algorithms. Before moving on to construct such algorithms, we summarize the main results in the following theorem, showing how our quantum algorithm is advantageous over the classical one. 

\begin{theorem}
    Provided the raw distances $d(\xbf_i, \xbf_j)$ between a pair $(\xbf_i,\xbf_j)$ among $N$ data points. Then:
    \begin{itemize}
    \item When $G$ is a tree, $W_1(x,y)$ can be estimated with accuracy $\epsilon$ using a quantum circuit of complexity
        $$ \mathcal{O}\left( \frac{1}{\epsilon}\log (N) \log^2\left( \frac{1}{\epsilon}\right) \kappa \log^2 \left( \frac{\kappa}{\epsilon} \right) + \frac{1}{\epsilon}\log (pq)\right). $$
        \item When $p=q$, $W_1(x,y)$ can be estimated with accuracy $\epsilon$ using a quantum circuit of complexity
        \begin{widetext}
            $$ \mathcal{O}\left(  \left(\log (N) \log^2\left( \frac{1}{\epsilon}\right) \kappa \log^2 \left(\frac{\kappa}{\epsilon}\right) + \log (p\log p)\right) \frac{1}{\gamma}p \log\left( \frac{p}{\epsilon}\right) \left( \frac{1}{ \epsilon} \right)\log^2 \frac{1}{\epsilon} \right),$$
        \end{widetext}
    \end{itemize}
    where in the above, $\kappa = \min_{i,j,q,k}\{ \frac{d_G(\xbf_i,\xbf_j)}{d_G(\xbf_k,\xbf_q)}   \} $, and $\gamma$ is a randomized factor that could be of $\mathcal{O}(1)$ in the best case and of $\mathcal{O}(\exp (p))$ in the worst case.  
\end{theorem}
The complexity of classical algorithms for the case $G$ being a tree and for the case $p=q$ is, respectively, $\mathcal{O}(p+q)$ and $\mathcal{O}(p^3)$. However, these complexities do not include the time required to compute the geodesic distances from the raw distances $d(\xbf_i, \xbf_j)$ between a pair $(\xbf_i,\xbf_j)$ among $N$ data points. Based on~\cite{hickok2023intrinsic}, the time complexity for producing geodesic distances from the raw distances is $\mathcal{O}\left( N^3\right)$. So the total classical complexity for estimating ORC are $\mathcal{O}\left( N^3 + pq \right)$ (for graph $G$ being a tree) and $\mathcal{O}\left(N^3  + p^3 \right) $ (for $p= q$), respectively. Compared to the quantum complexities above, there is exponential speed-up in the total number of points $N$ in both cases. Regarding $p,q$ (for the first case where $G$ is a tree), which controls the size of local neighborhood of two points $x,y$ of interest, there is still exponential speedup in the first case (when $G$ is a tree). However, in the second case, the degree of speedup depends on the factor $\gamma$. As $\gamma$ could be of $\mathcal{O}(1)$ in the best case and of $\mathcal{O}(\exp p)$ in the worst case, the best possible quantum speedup is polynomial. We remark one thing that as $p$ (and $q$) controls the size of local neighborhood of $x$ (and $y$), in principle, it can be chosen to be $\mathcal{O}(1)$. Thus, the scaling on $p$ is $\mathcal{O}(1)$ in all cases and therefore, our quantum algorithm can guarantee to provide significant speed-up with respect to $N$ -- the number of data points, which is typically large in reality.


\section{Quantum Algorithm}
\label{sec: quantumalgorithm}
Before describing our main quantum algorithm, we provide some preliminaries. Our subsequent proposal shall be utilizing the block-encoding and quantum singular value transformation (QSVT) framework~\cite{gilyen2019quantum}. This framework was built on the qubitization and quantum signal processing technique~\cite{low2017optimal,low2019hamiltonian}. In particular, it was shown in~\cite{gilyen2019quantum} that the block-encoding/QSVT framework can unify many known quantum algorithms, thus providing a universal language for this area, motivating many subsequent refinements~\cite{chakraborty2018power, nghiem2023improved, nghiem2025improved}. A key component of the block-encoding/QSVT framework is the unitary block-encoding of some operator. Specifically, a unitary $U$ is said to be an exact block-encoding of a matrix $A$ (with operator norm $||A||\leq 1$) if $A = ( \bra{\bf 0}\otimes \Ibb)  U (\Ibb \otimes \ket{0})$. If $ ( \bra{\bf 0}\otimes \Ibb)  U (\Ibb \otimes \ket{0}) = \tilde{A}$ and $||\tilde{A}-A|| \leq \epsilon$, then $U$ is said to be an $\epsilon$-approximated block-encoding of $A$. Given two unitary $U_1,U_2$ that block-encode $A_1,A_2$, then there is an efficient procedure that can construct the block-encoding of many operators, including linear combination of $A_1,A_2$, product of $A_1, A_2$, tensor product $A_1 \otimes A_2$, etc. A fundamental result of the QSVT~\cite{gilyen2019quantum} is that, given a unitary $U$ that block-encodes $A$, then the unitary block-encoding of $f(A)$ can be constructed efficiently, where $f(\cdot)$ is some computable function. We note that a more complete summary of block-encoding and related recipes can be found in the Appendix~\ref{sec: summaryofnecessarytechniques}.

Our quantum algorithm, described below, is a continuation of the proposal initiated in~\cite{nghiem2025quantum}, where the Gaussian curvature was considered. An important ingredient to our subsequent discussion is the following lemma:
\begin{lemma}[Appendix D of \cite{nghiem2025quantum}]
\label{lemma: diagonaldG}
    Provided the raw distances $d(\xbf_i, \xbf_j)$ between a pair $(\xbf_i,\xbf_j)$ among $N$ data points, then the $\epsilon$-approximated block-encoding of the following operator
\begin{align}
    \frac{1}{\alpha} \sum_{i,j=1}^N \ket{i-1}\bra{i-1}\otimes d^4_G(\xbf_i,\xbf_j) \ket{j-1}\bra{j-1}
    \label{eqn: dg4}
\end{align}
can be obtained by a quantum circuit of depth:
\begin{align}
    \mathcal{O}\left( \log (N) \log^2\left( \frac{1}{\epsilon}\right)  \right),
\end{align}
where $\alpha \in \mathcal{O}(1)$ is a constant that generally depend on the data set.
\end{lemma}
Moreover, we point out the following result of~\cite{chakraborty2018power} and~\cite{gilyen2019quantum}:
\begin{lemma}[Positive Power Exponent~\cite{gilyen2019quantum},\cite{chakraborty2018power}]
\label{lemma: positive}
    Given a block encoding of a positive matrix $\mathcal{M}/\gamma$ such that 
    $$ \frac{\Ibb}{\kappa_M} \leq \frac{\mathcal{M}}{\gamma} \leq \Ibb, $$
   and let $c \in (0,1)$, then we can implement an $\epsilon$-approximated block encoding of $(\mathcal{M}/\gamma)^c/2$ in time complexity $\mathcal{O}( \kappa_M T_M \log^2 (\frac{  \kappa_M}{\epsilon})  )$, where $T_M$ is the complexity to obtain the block encoding of $\mathcal{M}/\gamma$.  
\end{lemma}
A direct application of Lemma~\ref{lemma: diagonaldG} and the lemma above (with $ c= \frac{1}{4}$) allows us to obtain the following lemma:
\begin{lemma}
\label{lemma: key}
The $\epsilon$-approximated block-encoding of the following
\begin{align}
    \frac{1}{\alpha^{1/4}} \sum_{i,j=1}^N \ket{i-1}\bra{i-1}\otimes d_G(\xbf_i,\xbf_j) \ket{j-1}\bra{j-1}
\end{align}
can be obtained using a quantum circuit of depth:
\begin{align}
    \mathcal{O}\left( \log (N) \log^2\left( \frac{1}{\epsilon}\right) \kappa \log^2 \frac{\kappa}{\epsilon} \right),
\end{align}
where $\kappa$   is the condition number of the operator in Eqn.~\ref{eqn: dg4}, which can be seen to be $\kappa = \min_{i,j,q,k}\{ \frac{d_G(\xbf_i,\xbf_j)}{d_G(\xbf_k,\xbf_q)}   \} $. 
\end{lemma}

We have mentioned in the previous section that the closed-form solution to Eqn.~\ref{optimal_transport} generally does not exist. Therefore, in the following, we consider case by case and outline the corresponding quantum algorithm. For simplicity, we begin with the case when $G$ is a tree, for which a closed-form solution does exist.

\subsection{The case when G is a tree}
\label{sec: case1}
As mentioned, the solution to Eqn.~\ref{optimal_transport} in this case is given by:
\begin{align}
        W_1(x, y) = \frac{1}{p}\sum_{i=1}^pd_G(x_i, x) + d_G(x, y) + \frac{1}{q}\sum_{j=1}^qd_G(y, y_j).
\end{align}
In Section \ref{sec: classicalalgorithm}, we have used the notation $ \mathcal{N}_x\setminus \{y\} \equiv \{ x_1,x_2,...,x_p \}$ to denote the set of neighbor points of $x$. 
The key lemma~\ref{lemma: diagonaldG} uses $\xbf_i$ to denote the $i$-th data point in the given data set. To make things more appropriate, without loss of generalization, we promote the following index interchange $x_i \leftrightarrow \xbf_{i}$. For simplicity, the set $\mathcal{N}_x \setminus \{y\} \equiv   \{ x_1,x_2,...,x_p \}$ is then labeled as:
\begin{align}
      \{ x_1,x_2,...,x_p \} \equiv \{ \xbf_{1}, \xbf_{2}, ..., \xbf_{p} \}.
\end{align}
In a similar manner, the set $\mathcal{N}_y \setminus \{ x \} = \{ y_1, y_2, ..., y_q \}$ is labeled as:
\begin{align}
       \{ y_1, y_2, ..., y_q \} \equiv \{ \xbf_{p+1}, \xbf_{p+2}, ..., \xbf_{p+q} \}.
\end{align}
In this convention, the indexes of $x_i, y_i$ (for $1\leq i \leq p$) are $i, p+i$, respectively. In particular, we set $x$ to have index $i_x$ and $y$ to have index $i_y$. To estimate $W_1(x,y)$, we assume that the indexes of $x$ and $y$ are classically known, as well as the neighbors of $x$ and $y$, which are $\{i_1,i_2,...i_p\}$ and $\{j_1,j_2,...,j_q\}$, respectively. Denote the index of $x$ by $i_x$,  we note that the block-encoded operator 
$${\cal D} \equiv  \frac{1}{\alpha^{1/4}} \sum_{i=1}^N \sum_{j=1}^N \ket{i-1}\bra{i-1}\otimes d_G(\xbf_i,\xbf_j) \ket{j-1}\bra{j-1}$$
can be decomposed as: 
\begin{align}
\begin{split}
    & \frac{1}{\alpha^{1/4}} \sum_{\xbf_i,\xbf_j \notin \mathcal{N}_x\setminus y } \ket{i-1}\bra{i-1}\otimes d_G(\xbf_i,\xbf_j) \ket{j-1}\bra{j-1} + \\ &\frac{1}{\alpha^{1/4}} \sum_{\xbf_i \in \mathcal{N}_x \setminus y } \ket{i_x-1}\bra{i_x-1}\otimes d_G( x,\xbf_{i}) \ket{i-1}\bra{i-1}.
    \end{split}
\end{align}
Let $U_d$ denote the unitary block-encoding of the above operator. Provided the classical knowledge of indices $\{1,2,3,...,p\} $ and also of x, the following state can be prepared using a depth-1 circuit:
\begin{align}
    \ket{\phi} = \frac{1}{\sqrt{p+1}} \ket{i_x-1}\otimes\sum_{i=1}^p  \ket{i-1},
\end{align}
using a quantum circuit of depth $\mathcal{O}(\log p)$ and $\mathcal{O}(1)$ ancilla, based on the state preparation method \cite{mcardle2022quantum, zhang2022quantum}. If we take the unitary $U_d$ and apply it to the state $\ket{\bf 0}\ket{\phi}$, according to Definition \ref{def: blockencode}, we obtain the state:
\begin{widetext}
\begin{align}
\begin{split}
    U_d \ket{\bf 0}\ket{\phi} &= \ket{\bf 0} \Big( \frac{1}{\alpha^{1/4}} \sum_{\xbf_i,\xbf_j \neq \mathcal{N}_x \setminus y } \ket{i-1}\bra{i-1}\otimes d_G(\xbf_i,\xbf_j) \ket{j-1}\bra{j-1} + \\ &\frac{1}{\alpha^{1/4}} \sum_{ \xbf_i\in \mathcal{N}_x \setminus y  } \ket{i_x-1}\bra{i_x-1}\otimes d_G( x,\xbf_{i}) \ket{i-1}\bra{i-1} \Big) \frac{1}{\sqrt{p+1}} \sum_{i=1}^p \ket{i_x-1} \ket{i-1} + \ket{\rm Garbage},
\end{split}
\end{align}
\end{widetext}
where $\ket{\rm Garbage}$ is orthogonal to the first part. It can be seen that the above state is equivalent to:
\begin{align}
    \frac{1}{\alpha^{1/4} \sqrt{p+1}}\ket{\bf 0} \sum_{i=1}^p d_G(x, \xbf_{i})\ket{i_x-1}\ket{i-1} + \ket{\rm Garbage}.
\end{align}
Next, we consider the overlaps between the above state and $\ket{\bf 0}\ket{\phi}$ again:
\begin{widetext}
\begin{align}
\begin{split}
&\bra{\bf 0}\bra{\phi} \cdot \left(  \frac{1}{\alpha^{1/4} \sqrt{p+1}}\ket{\bf 0} \sum_{i=1}^p d_G(\xbf, \xbf_{i})\ket{i_x-1}\ket{i-1} + \ket{\rm Garbage}\right) \\ &=  \left(\frac{1}{\sqrt{p+1}} \sum_{i=1}^p \bra{i_x-1} \bra{i-1} \right)  \times \left(  \frac{1}{\alpha^{1/4} \sqrt{p+1}}\ket{\bf 0} \sum_{i=1}^p d_G(x, \xbf_{i})\ket{i_x-1}\ket{i-1} + \ket{\rm Garbage}\right) \\
&= \frac{1}{\alpha^{1/4} (p+1)} \sum_{i=1}^p d_G(x, \xbf_{i}) = \frac{1}{\alpha^{1/4} (p+1)} \sum_{i=1}^p d_G(x, x_i).
\end{split}
\end{align}    
\end{widetext}
These overlaps can be estimated via standard methods, such as the Hadamard test, SWAP test, or amplitude estimation. The value of the desired quantity $\frac{1}{p}\sum_{i=1}^pd_G(x_i, x)$ can be derived from the above estimation accordingly, e.g., by multiplying $\frac{1}{\alpha^{1/4} (p+1)} \sum_{i=1}^p d_G(x, x_i)$ with $ \frac{\alpha^{1/4} (p+1)}{p}$.

Repeating the above procedure but interchanging $x \leftrightarrow y, p \leftrightarrow q$, and $\{ x_1,x_2,...,x_p\} \leftrightarrow \{y_1,y_2,...,y_q\}$ can allow us to estimate the summation $\frac{1}{q}\sum_{j=1}^qd_G(y, y_j)$. The value $d_G(x,y)$ is, in fact, a special case of the above estimation,
as we just need to set a fixed value of index $x,y$. Instead of preparing $\ket{\phi}$ as above, we just need to prepare the state $\ket{i_x-1}\ket{i_y-1}$ and execute the same procedure. From the value of $W_1(x,y)$ and also $d_G(x,y)$, the estimation fo the Ollivier-Rici curvature is straightforward as $\kappa(x,y) = 1 - \frac{W_1(x,y)}{d_G(x,y)}  $.  \\

\noindent
\textbf{Complexity.} To analyze the total quantum circuit depth for this part, we summarize the above procedure in the following steps.
\begin{itemize}
    \item \textbf{Step 1.} Obtain the $\epsilon$-approximated unitary block-encoding $U_d$ of 
    \begin{align}
    \frac{1}{\alpha^{1/4}} \sum_{i=1}^N\sum_{j=1}^N \ket{i-1}\bra{i-1}\otimes d_G(\xbf_i,\xbf_j) \ket{j-1}\bra{j-1}
\end{align}
via Lemma \ref{lemma: key}, which involves a quantum circuit of depth:
\begin{align}
    \mathcal{O}\left( \log (N) \log^2\left( \frac{1}{\epsilon}\right) \kappa \log^2 \frac{\kappa}{\epsilon} \right).
\end{align}
     \item \textbf{Step 2.} The state $\ket{\phi}$ (and hence, $\ket{\bf 0}\ket{\phi}$) can be prepared using a circuit of depth $\mathcal{O}\left( \log p \right)$. Therefore, the quantum circuit depth that prepares the state $U_d \ket{\bf 0}\ket{\phi}$ is:
     \begin{align}
         \mathcal{O}\left( \log (N) \log^2\left( \frac{1}{\epsilon}\right) \kappa \log^2 \left( \frac{\kappa}{\epsilon} \right) + \log p \right).
     \end{align}
     
     \item \textbf{Step 3.} The overlap estimation with a precision $\epsilon$ incurs a total complexity 
     \begin{align}
        \mathcal{O}\left( \frac{1}{\epsilon}\log (N) \log^2\left( \frac{1}{\epsilon}\right) \kappa \log^2 \left( \frac{\kappa}{\epsilon} \right) + \frac{1}{\epsilon}\log p \right).
     \end{align}
     For completeness, we recall that $\kappa = \min_{i,j,q,k} \{ \frac{d_G(\xbf_i,\xbf_j)}{d_G(\xbf_k,\xbf_q)}   \}$. 
     \item \textbf{Step 4.} Finally repeat the above step with the interchange $x \leftrightarrow y, p \leftrightarrow q$, $\{ x_1,x_2,...,x_p\} \leftrightarrow \{y_1,y_2,...,y_q\}$, leads to the same complexity as above, but with $p$ replaced by $q$:
      \begin{align}
         \mathcal{O}\left( \frac{1}{\epsilon}\log (N) \log^2\left( \frac{1}{\epsilon}\right) \kappa \log^2 \left( \frac{\kappa}{\epsilon} \right) +\frac{1}{\epsilon} \log q \right).
     \end{align}
     Hence, the total complexity is:
     \begin{align}
         \mathcal{O}\left( \frac{1}{\epsilon}\log (N) \log^2\left( \frac{1}{\epsilon}\right) \kappa \log^2 \left( \frac{\kappa}{\epsilon} \right) +\frac{1}{\epsilon} \log (pq) \right).
     \end{align}
\end{itemize}

\subsection{The case when $p=q $}
\label{sec: case2}
This is considerably harder than in the previous case, as we have discussed in Section \ref{sec: classicalalgorithm}.  Although the Hungarian algorithm \cite{kuhn1955hungarian} can solve this case exactly in polynomial time, below, we shall attempt to derive a quantum algorithm running in linear time, at the cost of achieving the approximation. We first recall that the reasoning of \cite{kuhn1955hungarian} results in the matrix $\gamma \equiv [ \gamma_{ij}]_{p \times p}$ being a permutation matrix multiplied by a factor of $\frac{1}{p}$. Given that the Earth Mover (1-Wasserstein) distance $W_1(x,y)$ is defined as
\begin{align}
     W_1(x,y) &= \min_{\gamma_{ij}}  \sum^p_{i=1}\sum^q_{j=1} \frac{1}{p}d_G(x_i,y_j) \gamma'_{ij}. 
\end{align}
where $\gamma'  \equiv [\gamma'_{ij}]$ is the permutation matrix. Then if we organize $\{ d_G(x_i,y_j)\}_{i,j=1}^p$ as a $p \times p$ matrix, and define $d_G \equiv [ d_G(x_i,y_j)]_{i,j=1}^p\equiv [ d_{ij}]_{i,j=1}^p$ be the $p \times p$ matrix with entries being the geodesic distances as indicated. The minimization problem can then be translated into the following problem: 
\begin{center}
    \textit{For each column index $i$, find an element $j$ so that their summation $ \sum_{i=1}^p d_{\pi(i),i}$ is minimized. In other words, we need to find the minimum value of $ \{  d_{i_11} + d_{i_22} + \cdots + d_{i_p p}  \}_{i_1 \neq i_2 \neq ... \neq i_p}^p $. }
\end{center}
How do we compute the above sum and minimize over all possible permutations?
Our algorithm is built on the following observation. Suppose, for now, that $p=3$. Then we have:
\begin{align}
    d_G = \begin{pmatrix}
        d_{11} & d_{12} & d_{13} \\
        d_{21} & d_{22} & d_{23} \\
        d_{31} & d_{32} & d_{33}
    \end{pmatrix}.
\end{align}
Consider the following $9\times 9$ diagonal matrix:
\begin{equation}
    D_G = \text{diag}\left( d_{11}, d_{21}, d_{31}, d_{12}, d_{22}, d_{32}, d_{31}, d_{32}, d_{33} \right).
\end{equation}
We define three blocks $D_1, D_2, D_3$ as follows:
\begin{align}
\begin{split}
    D_1 &= \text{diag}\left(d_{11}, d_{21}, d_{31} \right), \\
    D_2 &= \text{diag}\left(d_{12}, d_{22}, d_{32} \right), \\
    D_3 &= \text{diag}\left(d_{13}, d_{23}, d_{33} \right).
\end{split}
\end{align}
Next, we consider the top-left corner sub-matrix of dimension $9 \times 9$, denoted by $[\cdot]_{9\times 9}$ of:  
\begin{align}
\begin{split}
    &[D_1 \otimes \Ibb_9]_{9\times 9} =
    \\
    & \ \ \ \ \text{diag}\left(d_{11}, d_{11}, d_{11}, d_{11}, d_{11}, d_{11}, d_{11}, d_{11}, d_{11} \right),
    \\
    &[\Ibb_3 \otimes D_2 \otimes \Ibb_3]_{ 9 \times 9} =
    \\
    & \ \ \ \ 
    \text{diag}\left(d_{12}, d_{12}, d_{12}, d_{22}, d_{22}, d_{22}, d_{32}, d_{32}, d_{32} \right),
    \\
    & [\Ibb_9  \otimes D_3]_{9 \times 9} = 
    \\
    & \ \ \ \ \text{diag}\left(  d_{13},d_{23},d_{33},d_{13},d_{23},d_{33},d_{13},d_{23},d_{33} \right).
\end{split}
\end{align}
Their summation is as follows (again, we pay attention to the top-left corner of dimension $9 \times 9$): 
\begin{equation}
    \begin{split}
        &[ D_1 \otimes \Ibb_9 + \Ibb_3 \otimes D_2 \otimes \Ibb_3 +  \Ibb_9  \otimes D_3 ]_{9\times 9}=
        \\
        &\ \text{diag}(d_{11}+ d_{12} +  d_{13}, d_{11} + d_{12}  +  d_{23}, d_{11} + d_{12} +  d_{33},
        \\
        & \ \ \ \ \ d_{11} + d_{22} + d_{13}, d_{11} +d_{22} +d_{23}, d_{11}+ d_{22} + d_{33}, 
        \\
        & \ \ \ \ \ d_{11}+ d_{32} + d_{13}, d_{11}+ d_{32}+ d_{23}, d_{11} + d_{32} + d_{33}).
    \end{split}
\end{equation}
We recall that the above operator is the top-left corner submatrix of $ D_1 \otimes \Ibb_9 + \Ibb_3 \otimes D_2 \otimes \Ibb_3 +  \Ibb_9  \otimes D_3 $ with dimension $9 \times 9$. If we pay attention to the next block submatrix of the same dimension $9 \times 9$, e.g., the submatrix whose row and column indices run from $9 \longrightarrow 18$ of matrix $ D_1 \otimes \Ibb_9 + \Ibb_3 \otimes D_2 \otimes \Ibb_3 +  \Ibb_9  \otimes D_3 $, then the structure is same as above but the entries $d_{11}$ is replaced by $d_{12}$. In a similar manner, the next submatrix of dimension $ 9 \times 9$, or the submatrix whose row and column indexes run from $18 \longrightarrow 27$ of $ D_1 \otimes \Ibb_9 + \Ibb_3 \otimes D_2 \otimes \Ibb_3 +  \Ibb_9  \otimes D_3$ has the diagonal entries being similar as above, but $d_{11}$ is replaced with $d_{13}$. What we can conclude from this observation is that
\begin{center}
    \textit{The matrix $ D_1 \otimes \Ibb_9 + \Ibb_3 \otimes D_2 \otimes \Ibb_3 +  \Ibb_9  \otimes D_3$ has the entries $\{  ( d_{i1} + d_{j2} + d_{k3} )\}^3_{ \substack{i,j,k=1 }}$ in the diagonal}.
\end{center}
By generalizing the above procedure to arbitrary value of $p$, we then have the following property, stated as a lemma for subsequent reference:
\begin{lemma}
\label{lemma: summationdiagonal}
Let $d_G(x_i,y_j)$ be the geodesic distances between the points $x_i$ and $y_j$ from the given dataset. Define $d_G \equiv [ d_G(x_i,y_j)]_{i,j=1}^p\equiv [ d_{ij}]_{i,j=1}^p $ as the corresponding geodesic distance matrix of those points belonging to the neighbor of two points $x,y$ of consideration. 

Let $D_G$ be the matrix of dimension $p \times p$ defined as $D_G \equiv \sum_{i,j=1}^p d_{ji} \ket{i} \bra{i} \otimes \ket{j}\bra{j} $ which contains the geodesic distances on the diagonal. Define $p$ block submatrices as $D_i = \bigoplus_{j=1}^p d_{ji} \ket{j}\bra{j}$ for $i=1,2,...,p$. Then it holds that
$$ \mathcal{D}_P\equiv D_1 \otimes \Ibb_{p }^{\otimes p-1} + \Ibb_{p} \otimes D_2 \otimes \Ibb_{p}^{\otimes p-2} + ... + \Ibb_{p}^{\otimes p-1} \otimes D_p $$
is diagonal and contains $ \{  d_{i_11} + d_{i_22} + \cdots + d_{i_p p}  \}_{i_1,i_2,..,i_p =1}^p $ in the diagonal. 
\end{lemma}

{However, we need to restrict to the cases where the indices $i_k$'s are all different. Namely, for the case ($p=q$) the objective reduces to finding the minimum of $ \{ \sum_{i=1}^p d_{\pi(i),i} \}_{i=1}^p $. The permutation requires that only a subset of $ \{  d_{i_11} + d_{i_22} + \cdots + d_{i_p p}  \}_{i_1,i_2,..,i_p =1}^p $ is of interest. In order to filter out those elements that correspond to the permutation of the indices, we shall construct a projector $\Pi$, see Eq.~(\ref{eq:ProjectorPi}) below, that achieves this goal.} Let us first point out the following property. Within the matrix $\mathcal{D}_P$, which has size $p^p \times p^p$, the entry having value $ d_{i_11} + d_{i_22} + \cdots + d_{i_p p} $ would have the location index (along the diagonal) as:
\begin{align}
    k  = (i_1-1)p^{p-1} + (i_2-1)p^{p-2} + \cdots + i_p. 
\end{align}
Because the summation $ \sum_{i=1}^p d_{\pi(i),i} = $ only contains the permutation for all $i$, it implies that the indices $i_1,i_2,...,i_p$ in the summation $ d_{i_11} + d_{i_22} + \cdots + d_{i_p p}$ must be different (no pair of indices having the same value). So, among all the diagonal entries, only those entries with the index $k  = (i_1-1)p^{p-1} + (i_2-1)p^{p-2} + \cdots + i_p  $ with $i_1 \neq i_2 \neq ... \neq i_p$ are of interest. To obtain these elements, we consider the projector \begin{equation}
    \Pi=\sum_{k:i_1 \neq i_2 \neq ... \neq i_p } \ket{k}\bra{k},
    \label{eq:ProjectorPi}
\end{equation}
whose construction is shown below right above Lemma~\ref{lemma: improveddme}. 
The product of this projector with $\mathcal{D}_P$ contains those  $\{ \sum_{i=1}^p d_{\pi(i),i} \}_{i=1}^p$ of interest.

Our problem then is to find the minimum diagonal entry of the matrix $\Pi  \mathcal{D}_P$. This is trivially equivalent to the minimum (nonzero) eigenvalue of the matrix. However, before proceeding with the eigenvalue solver, we describe a procedure for constructing the block-encoding of $D_G$, as well as that of $\mathcal{D}_P, \Pi \mathcal{D}_P$.

\smallskip
\noindent \textit{\underline{Block encoding of $D_G$.}} First,
from Lemma~\ref{lemma: key}, we have the block-encoding of 
\begin{align}
\begin{split}
   & \frac{1}{\alpha^{1/4}} \sum_{i=1}^N \sum_{j=1}^N \ket{i-1}\bra{i-1}\otimes d_G(\xbf_i,\xbf_j) \ket{j-1}\bra{j-1},
   \end{split}
\end{align}
which contains the geodesic distances between \textit{every pair} of points in the given data. Our goal now is to first obtain the (block-encoding of) operator $D_G$ that appears in the lemma above, which only includes certain data points of interest, before finding the minimum of $ D_1 \otimes \Ibb_{p }^{\otimes p} + \Ibb_{p} \otimes D_2 \otimes \Ibb_{p}^{\otimes p-1} + ... + \Ibb_{p}^{\otimes p} \otimes D_p$.

In the previous section, we have defined the classical indexes $\{\xbf_1,\xbf_2, ..., \xbf_p\}$ and $\{\xbf_{p+1}, \xbf_{p+2}, ..., \xbf_{p+q} \}$ of the neighbors of $x$ and $y$, respectively. Now we define the permutation operator $U_p$ of dimension $N^2 \times N^2$ that acts as follows:
\begin{align}
\begin{cases}
    U_p \ket{i} \ket{j } \leftrightarrow  \ket{i}\ket{j-p}\\
    \text{arbitrary for the remaining basis}
\end{cases}
\end{align}
An arbitrary permutation of a given dimension can be constructed with polylogarithmic resources~\cite{shende2005synthesis, barenco1995elementary}. That being said, the unitary $U_p$ can be constructed using a quantum circuit of depth $\mathcal{O}\left( \log (N \log N )\right)$. However, we note that we are only limiting ourselves to the space spanned by those neighbors of $x$ and $y$ (with the corresponding indexes $\{ 1,2,3,...,p, p+1,p+2, ..., p+p\}$, which is effectively of dimension $p^2$. Thus, the unitary $U_p$ can be effectively realized with a circuit of depth $\mathcal{O}\left( \log (p \log p) \right)$. 

Given the unitary $U_p$, according to Def.~\ref{def: blockencode}, it block-encodes itself. Then we use Lemma \ref{lemma: product} to construct the block-encoding of:
\begin{align}
\begin{split}
     & U_p \frac{1}{\alpha^{1/4}} \sum_{i=1}^N\sum_{j=1}^N \ket{i-1}\bra{i-1}\otimes d_G(\xbf_i,\xbf_j) \ket{j-1}\bra{j-1} U_p^\dagger \\
     &= \frac{1}{\alpha^{1/4}} \sum_{i,j=1}^{p} d_G(\xbf_i,\textbf{y}_j) \ket{i-1}\bra{i-1}\otimes \ket{j-1}\bra{j-1} + (...),  \\
      &= \frac{1}{\alpha^{1/4}} \sum_{i,j=1}^{p} d_{ij} \ket{i-1}\bra{i-1}\otimes \ket{j-1}\bra{j-1} + (...),  
\end{split}
\end{align}
where $(...)$ refers to the unwanted part (which contains the distance between other pairs), and we only pay attention to the top-left submatrix of size $p\times p$.  

Next, we define the SWAP operator that performs the swapping between two sub-systems of dimension $p$, e.g., $ \ket{i-1}\bra{i-1} \leftrightarrow \ket{j-1}\bra{j-1} $. This operator can be realized with a circuit of depth $\mathcal{O}\left( \log p \right)$ \cite{shende2005synthesis} (in fact, they are a specific kind of permutation). We then use Lemma~\ref{lemma: product} to construct the block-encoding of: 
\begin{align}
\begin{split}
   & \text{SWAP} \cdot \frac{1}{\alpha^{1/4}} \sum_{i=1}^{p} \sum_{j=1}^{p} d_{ij} \ket{i-1}\bra{i-1}\otimes \ket{j-1}\bra{j-1} {\rm SWAP}^\dagger   \\
    &  = \frac{1}{\alpha^{1/4}} \sum_{i=1}^{p} \sum_{j=1}^{p} d_{ij}  \ket{j-1}\bra{j-1} \otimes \ket{i-1}\bra{i-1}. 
    \end{split}
\end{align}
 Up to the scaling factor $\alpha^{1/4}$, the above operator contains the geodesic distances $\{ d_{ij} \equiv d_G(\xbf_{i}, \textbf{y}_{j}) \}$ on the diagonal.  Defining the above operator multiplied by $\alpha^{1/4}$ as $D_G$, therefore, we have obtained the block-encoding of $D_G/\alpha^{1/4}$. 

\smallskip
\noindent
\textit{\underline{Block-encoding of $\mathcal{D}_P$.}} We now discuss how to construct the  block encoding of $\mathcal{D}_P\equiv D_1 \otimes \Ibb_{p }^{\otimes p} + \Ibb_{p} \otimes D_2 \otimes \Ibb_{p}^{\otimes p-1} + ... + \Ibb_{p}^{\otimes p} \otimes D_p$. First, we note that $D_G = \sum_{i=1}^p  \ket{i}\bra{i} \otimes D_i$. Next, we define $p$ different permutations $U_{1}, ..., U_{p}$ that act as follows: for any $i=1,2,...p$,
\begin{align}
    U_i \ket{i } \leftrightarrow  \ket{0}
\end{align}
and the remaining bases are permuted arbitrarily (for all $i$). Based on~\cite{shende2005synthesis, barenco1995elementary}, these unitaries can be constructed using circuits of depth $\mathcal{O}\left( \log (p\log p) \right)$. Next, for each $i$, we use $U_i$ which block-encodes itself, and Lemma~\ref{lemma: product} to construct the block-encoding of:
\begin{align}
    U_i \frac{D_G}{\alpha^{1/4}} U_i^\dagger &= U_i  \frac{ \sum_{i=1}^p  \ket{i}\bra{i} \otimes D_i}{\alpha^{1/4}} U_i^\dagger \\
    &= \frac{1}{\alpha^{1/4}}\ket{0}\bra{0}\otimes D_i  + (...),
\end{align}
where $(...)$ refers to the unwanted part. It can be seen that, from the Definition~\ref{def: blockencode}, the above operator is exactly the block-encoding of $\frac{1}{\alpha^{1/4}}D_i$. Repeating the same procedure for different permutations $U_i$'s, we obtain the block-encoding of all $\frac{1}{\alpha^{1/4}}D_i $ for $i=1,2,...,p$. 

The remaining task for constructing $\mathcal{D}_P$ is straightforward as follows. As the block encoding for the identity matrix of arbitrary dimension can be constructed easily (see Def.~\ref{def: blockencode}), we can employ Lemma \ref{lemma: tensorproduct} to construct the block-encoding of the following operators
\begin{align}
    \frac{1}{\alpha^{1/4}}D_1 \otimes \Ibb_{p }^{\otimes p-1},  \frac{1}{\alpha^{1/4}}\Ibb_{p} \otimes D_2 \otimes \Ibb_{p}^{\otimes p-2} , ... , \frac{1}{\alpha^{1/4}}\Ibb_{p}^{\otimes p-1} \otimes D_p.
\end{align}
Then we use Lemma~\ref{lemma: sumencoding} to construct the block encoding of their linear combination,  
\begin{widetext}
    \begin{align}
     &\frac{1}{p}\left( \frac{1}{\alpha^{1/4}}D_1 \otimes \Ibb_{p }^{\otimes p-1} +  \frac{1}{\alpha^{1/4}}\Ibb_{p} \otimes D_2 \otimes \Ibb_{p}^{\otimes p-2} + ... + \frac{1}{\alpha^{1/4}}\Ibb_{p}^{\otimes p-1} \otimes D_p \right) \\
     &=\frac{1}{p\alpha^{1/4}} \left( D_1 \otimes \Ibb_{p }^{\otimes p-1} + \Ibb_{p} \otimes D_2 \otimes \Ibb_{p}^{\otimes p-2} + ... + \Ibb_{p}^{\otimes p-1} \otimes D_p \right)=\frac{1}{p\alpha^{1/4}}\mathcal{D}_P. 
\end{align}
\end{widetext}

\smallskip
\noindent
\textit{\underline{Block-encoding of $\Pi$. }}
\textcolor{black}{
Next, we show how to obtain the block-encoding of the projector $\Pi$ in Eq.~(\ref{eq:ProjectorPi})  that allows us to restrict the indices to the permutations only, as remarked earlier.  We first consider the state $\frac{1}{ \sqrt{p!}} \sum_{k:i_1 \neq i_2 \neq ... \neq i_p } \ket{k}$, which can be prepared using any of the state preparation methods in~\cite{zhang2022quantum, marin2023quantum, nakaji2022approximate, zoufal2019quantum}. These methods generally have circuit complexity $\mathcal{O}\left( \log (p!)\right)$. However, the method in~\cite{zhang2022quantum} would requires $\mathcal{O}(p!)$ ancilla qubit, meanwhile the methods in~\cite{marin2023quantum, nakaji2022approximate} requires $\mathcal{O}(1)$ ancilla qubits; see e.g., Theorem 1 of~\cite{marin2023quantum}. In practice the value of $p=\mathcal{O}(1)$, therefore practically either of these methods can be used. However, we prefer the methods in~\cite{marin2023quantum, nakaji2022approximate} as it requires less ancilla qubits. Given the preparation of such the state, then we append the ancilla initialized in $\ket{0}^{\otimes \log p^p}$, and use CNOT gates to transform $\frac{1}{ \sqrt{p!}} \sum_{k:i_1 \neq i_2 \neq ... \neq i_p } \ket{k} \ket{0}^{\otimes \log p^p} $   to $\frac{1}{ \sqrt{p!}} \sum_{k:i_1 \neq i_2 \neq ... \neq i_p } \ket{k}\ket{k}$. Tracing out either of these register yields the normalized projector $\frac{1}{p!}\Pi$, which can be block-encoded via the following lemma:
\begin{lemma}[Block encoding of density matrix, see e.g.~\cite{gilyen2019quantum}]
\label{lemma: improveddme}
Let $\rho = \Tr_A \ket{\Phi}\bra{\Phi}$, where $\rho \in \mathbb{H}_B$ and $\ket{\Phi} \in  \mathbb{H}_A \otimes \mathbb{H}_B$. Given a unitary $U$ that prepares $\ket{\Phi}$ from $\ket{\bf 0}_A \otimes \ket{\bf 0}_B$, there exists a highly efficient procedure that constructs an exact unitary block encoding of $\rho$ using $U$ and $U^\dagger$ once each.
\end{lemma}}

\smallskip
\noindent
\textit{\underline{Block-encoding of $\Pi \mathcal{D}_P$.}} From the block-encoding of $\frac{1}{p!}\Pi $ and of $\frac{1}{p\alpha^{1/4}}\mathcal{D}_P $, we can use Lemma~\ref{lemma: product} to construct the block-encoding of $ \frac{1}{p!p\alpha^{1/4}} \Pi \mathcal{D}_P. $

\medskip
\noindent
\textit{\underline{Finding the Earth Mover distance.} } Earlier, we have mentioned that the operator $\Pi \mathcal{D}_P$ contains the summations $ \{ \sum_{i=1}^p d_{\pi(i),i} \}_{i=1}^p$ of interest along the diagonal. It can be seen that for a diagonal matrix, the minimum entry along the diagonal is also the minimum eigenvalue. Therefore, our challenge now is to find the minimum eigenvalue of this block-encoded operator $\frac{1}{(p+1)!\alpha^{1/4}} \Pi \mathcal{D}_P $, which contains the desired minimum value of the Earth Mover distance $W_1(x,y)$ (up to a scaling). To this end, we point out the following result of \cite{nghiem2023improved, nghiem2025improved}, which is based on the classical power method for finding the largest eigenvalues/eigenvectors:
\begin{lemma}
\label{lemma: powermethod}
    Let $A \in \Rbb^{n \times n}$ be a Hermitian matrix of operator norm $||A||_{\rm operator} \leq 1$. Let $U_A$ be the block-encoding of $A$ with circuit complexity $T_A$. Then its largest eigenvalue $\psi_{\max}$ can be estimated up to an additive error $\epsilon$ using a quantum circuit of depth
    $$\mathcal{O}\left( T_A \left( \frac{1}{\Delta \gamma \epsilon} \right) \log \left(\frac{n}{\epsilon} \right)\log \left( \frac{1}{\epsilon}\right) \right),$$
    where $\Delta$ is the gap between the largest and the second largest eigenvalue. $\gamma$ is defined as $|\braket{\psi_0, \psi_{\max}}|$ where $ \ket{\psi_{\max}}$ is the eigenstate corresponding to the largest eigenvalue, and $\ket{\psi_0}$ is some randomly initialized state. 
\end{lemma}
While the above lemma aims at finding the largest eigenvalues/eigenvectors, it can be adapted to deal with the minimum ones by considering the matrix $A'$ to be the inverse of $A$ or the pseudoinverse of $A$ if $A$ is singular. In either case, the maximum nonzero eigenvalue of $ A'$ is equal to $\frac{1}{\psi_{\min}}$. An estimation of the largest eigenvalue of $A'$ can thus be used to infer the minimum eigenvalue $\psi_{\min}$ of $A$. The block-encoding of $\frac{A^{-1}}{\kappa_A}$ (where $\kappa_A$ is the condition number of $A$) can be obtained  from $A$ by using matrix inversion technique of \cite{childs2017quantum}, or QSVT with a proper polynomial approximation to $A^{-1}$, e.g., see Corollary 67 in \cite{gilyen2019quantum} and the discussion below such the corollary. As the inversion incurs a circuit complexity by $\mathcal{O}\left( \kappa_A \log \frac{1}{\epsilon} \right)$, so the total complexity for using the above lemma to find the minimum eigenvalue of $A$ is 
$$\mathcal{O}\left( T_A \kappa_A \left( \frac{1}{\Delta \gamma \epsilon} \right) \log \left(\frac{n}{\epsilon} \right)\log^2 \left( \frac{1}{\epsilon}\right) \right).$$

The application of the above discussion to our context is straightforward to estimate:
\begin{align}
   \min  \frac{1}{p!p \alpha^{1/4}} \{  \sum_{i=1}^p d_{\pi(i),i} \}_{i=1}^p \equiv \frac{W_1(x,y)}{p!\alpha^{1/4}}.
\end{align}
from which the value of $ W_1(x,y)$ can be inferred by multiplying such estimation with $ p!\alpha^{1/4}$.  

\bigskip
\noindent
\textbf{Complexity.} To analyze the complexity, we recapitulate the algorithm above as follows. 
\begin{itemize}
    \item \textbf{Step 1.} Obtain the block-encoding of 
    \begin{align}
    \frac{1}{\alpha^{1/4}} \sum_{i,j=1}^N \ket{i-1}\bra{i-1}\otimes d_G(\xbf_i,\xbf_j) \ket{j-1}\bra{j-1}
\end{align}
via Lemma \ref{lemma: key}, which involves a quantum circuit of depth:
\begin{align}
    \mathcal{O}\left( \log (N) \log^2\left( \frac{1}{\epsilon}\right) \kappa \log^2 \frac{\kappa}{\epsilon} \right).
\end{align}
\item \textbf{Step 2.} From the block-encoding of the above operator, construct the block-encoding of 
$$ \frac{1}{\alpha^{1/4}} \sum_{m,k=1}^{p} d_{mk} \ket{m-1}\bra{m-1}\otimes \ket{k-1}\bra{k-1} + (...),$$
which uses further a permutation operator and a SWAP operator of dimension $p \times p$, with circuit depth $\mathcal{O}\left( \log (p\log p)\right)$. So the total complexity is:
\begin{align}
    \mathcal{O}\left( \log (N) \log^2\left( \frac{1}{\epsilon}\right) \kappa \log^2 \frac{\kappa}{\epsilon} + \log (p\log p)\right).
\end{align}
\item \textbf{Step 3.} Use permutation operator of dimension $p \times p$ again to construct the block-encoding of $ \frac{D_i}{\alpha^{1/4}}$ for all $i=1,2,...,p$. The total complexity at this point is 
\begin{align}
    \mathcal{O}\left( \log (N) \log^2\left( \frac{1}{\epsilon}\right) \kappa \log^2 \frac{\kappa}{\epsilon} + \log (p\log p)\right).
\end{align}
\item \textbf{Step 4.} Use Lemma \ref{lemma: sumencoding} to construct the block-encoding of
$$ A\equiv  \frac{ \left( D_1 \otimes \Ibb_{p }^{\otimes p-1} + \Ibb_{p} \otimes D_2 \otimes \Ibb_{p}^{\otimes p-2} + ... + \Ibb_{p}^{\otimes p-1} \otimes D_p \right)}{p\alpha^{1/4}},$$
which incurs a total quantum circuit depth
\begin{align}
    \mathcal{O}\left( p\log (N) \log^2\left( \frac{1}{\epsilon}\right) \kappa \log^2 \frac{\kappa}{\epsilon} + p\log (p\log p)\right).
\end{align}
\item \textbf{Step 5.} Use the modified version of Lemma~\ref{lemma: powermethod} with $A$ being defined as above to find the minimum eigenvalue, which can be used to infer the desired value $ W_1(x,y)$ (up to the scaling $p!\alpha^{1/4})$, resulting in the overall complexity:
\begin{widetext}
\begin{align}
    \mathcal{O}\left(  \left(\log (N) \log^2\left( \frac{1}{\epsilon}\right) \kappa \log^2 \left(\frac{\kappa}{\epsilon}\right) + \log (p\log p)\right) p \log\left( \frac{p}{\epsilon}\right) \left( \frac{1}{\Delta \gamma \epsilon} \right)\log^2 \frac{1}{\epsilon} \right), 
\end{align}
\end{widetext}
where in this case $\Delta$ is defined as the gap between the smallest and second smallest eigenvalue of the operator $A$; and we recall that $\kappa = \min_{i,j,q,k}\{ \frac{d_G(\xbf_i,\xbf_j)}{d_G(\xbf_k,\xbf_q)}   \} $. 
\end{itemize}

\noindent
\textbf{Discussion.} As indicated above, the complexity depends on a parameter $\gamma$, which is defined as  $|\braket{\psi_0, \psi_{\min}}|$  where $\ket{\psi_0}$ is the initially randomized vector, and $\ket{\psi_{\min}}$ is the eigenstate corresponding to the minimum eigenvalue of $A$ above. As it involves a randomization, the value of $\gamma$ also behaves probabilistically. If we are lucky to have large overlaps, then $\gamma$ is of $\mathcal{O}(1)$. Otherwise, in the worst case scenario, $\gamma$ can be exponentially small as the dimension of matrix, and thus the complexity could scale as $\mathcal{O}(p^p)$, which is exponential in $p$. Therefore, in order for our method maintains its efficiency, the value of $p$ should be small, of order $\mathcal{O}(1)$. This can be achieved by choosing a small neighborhood around $x $ and $y$. 

Earlier, we have noted that our definition of Earth Mover distance does not include the point $x,y$, and that adding these points (according to \cite{saucan2019discrete,math8091416}) would not change the procedure as well as the complexity of our algorithm.

\section{Conclusion}
\label{sec: conclusion}
In this work, we have constructed a quantum algorithm for estimating the Ollivier-Ricci curvature in two specific classes of problems, an important quantity with practical applications. Our work incorporated a wide range of methods, including optimal transport theory (specifically its discrete version), linear programming, and, notably, the block-encoding/QSVT framework introduced in~\cite{gilyen2019quantum}. By leveraging the recent result in~\cite{nghiem2025quantum}, we are able to obtain the block-encoding of an operator that contains the geodesic distances between data points as diagonal entries. From such a (block-encoded) operator, we showed that a series of QSVT techniques can be applied appropriately, resulting in the estimation of the Earth Mover (1-Wasserstein) distance $W_1(x,y)$ for two classes of the problem:  (1)  when $G$ forms a tree and (2) the case $p=q$, i.e., number of the respective neighbors of two points on an edge $(x,y)$ is the same. Estimating the Ollivier-Ricci curvature in these cases can be straightforwardly done by quantum methods. We then provide an analysis showing that our quantum algorithm admits logarithmic running time in the number of data points, and also the (local) size of the neighborhood of interest. Thus, our quantum algorithm offers an exponential speedup compared to the classical algorithm. Our result has strongly highlighted the potential of quantum computation for geometrical data analysis, adding another example to our previous attempt~\cite{nghiem2025quantum}. What other problems in which our technique can be proved useful is thus of high interest. For example, the arbitrary case of the Ollivier-Ricci curvature remains open and is left for future exploration.

\section*{Acknowledgements}
This work was supported by the U.S. Department of Energy, Office of Science, National Quantum Information Science Research Centers, Co-design Center for Quantum Advantage (C2QA) under Contract No. DE-SC0012704. N.A.N. and T.-C.W also acknowledge support from the Center for Distributed Quantum Processing at Stony Brook University. N.A.N. thanks the hospitality of Harvard University where he has an academic visit during the completion of this project.

\bibliography{ref.bib}
\bibliographystyle{unsrt}

\appendix
\onecolumngrid

\section{Calculating ORC \label{app:ORC_calc}}

Here we show the concrete steps required to estimate the ORC associated with edge $(x,y)$ of the graph in Fig. \ref{fig:orc}. The geodesic distances between pairs of relevant points for this calculation are:
\begin{equation}
    \begin{split}
        d_G(x_1,y_1) = 1 \ \ , \ \ d_G(x_1,y_2) = 3 \ \ , \ \ d_G(x_1,y_3) = 3 \ \ , \ \ d_G(x_1,y_4) = 2 \ \ ,  
        \\
        d_G(x_2,y_1) = 2 \ \ , \ \ d_G(x_2,y_2) = 3 \ \ , \ \ d_G(x_2,y_3) = 3 \ \ , \ \ d_G(x_2,y_4) = 3 \ \ , 
        \\
        d_G(x_3,y_1) = 3 \ \ , \ \ d_G(x_3,y_2) = 2 \ \ , \ \ d_G(x_3,y_3) = 2 \ \ , \ \ d_G(x_3,y_4) = 3 \ \ . 
    \end{split}
\end{equation}
We need to find the non-negative $3 \times 4 = 12$ values $\{ \gamma_{ij}\}_{i=1,j=1}^{3,4}$ that minimize the following linear combination:
\begin{equation}
\begin{split}
    \mathcal{W}(\{ \gamma_{ij}\}) = \ & d_G(x_1,y_1) \gamma_{11} + d_G(x_1,y_2) \gamma_{12} + d_G(x_1,y_3) \gamma_{13} + d_G(x_1,y_4) \gamma_{14}  \ + 
    \\
    &d_G(x_2,y_1) \gamma_{21} + d_G(x_2,y_2) \gamma_{22} + d_G(x_2,y_3) \gamma_{23} + d_G(x_2,y_4) \gamma_{24}  \ +
    \\
    &d_G(x_3,y_1) \gamma_{31} + d_G(x_3,y_2) \gamma_{32} + d_G(x_2,y_3) \gamma_{33} + d_G(x_3,y_4) \gamma_{34} 
    \\
    = \ &\gamma_{11} + 3\gamma_{12} + 3\gamma_{13} + 2\gamma_{14} + 
    2\gamma_{21} + 3\gamma_{22} +
    \\
    & 3\gamma_{23} + 3\gamma_{24} + 3\gamma_{31} + 2 \gamma_{32} + 2 \gamma_{33} + 3\gamma_{34} \ ,
\end{split}
\end{equation}
subjected to the constraints:
\begin{equation}
\begin{split}
    \gamma_{11} + \gamma_{12} + \gamma_{13} + \gamma_{14} & 
    \\
  =   \gamma_{21} + \gamma_{22} + \gamma_{23} + \gamma_{24} & 
    \\
 =    \gamma_{31} + \gamma_{32} + \gamma_{33} + \gamma_{34} & = 1/3
\end{split}
\end{equation}
and
\begin{equation}
\begin{split}
    \gamma_{11} + \gamma_{21} + \gamma_{31} & 
    \\
  =   \gamma_{12} + \gamma_{22} + \gamma_{32} & 
    \\
  =   \gamma_{13} + \gamma_{23} + \gamma_{33} &
    \\
  =    \gamma_{14} + \gamma_{24} + \gamma_{34} & = 1/4 \ .
\end{split}
\end{equation}
The minimum value of $\mathcal{W}$ can be found to be $25/12$, located at 
$$\gamma_{11}=1/4 \ , \ \gamma_{14}=1/12 \ , \ \gamma_{23} = 1/6 \ , \ \gamma_{24} = 1/6 \ , \ \gamma_{32} = 1/4 \ , \ \gamma_{33} = 1/12 \ , $$
and other $\gamma$-values are $0$. From this result, we can use Eq. \eqref{optimal_transport} to estimate the Earth Mover distance 
$$W_1(x,y) =  \min_{\{\gamma_{ij}\}} \mathcal{W}(\{\gamma_{ij}\}) =  25/12   . $$
The geodesic distance between point $x$ and point $y$ is of unit-length, i.e. $d_G(x,y)=1$, therefore we can obtain the estimated ORC value from Eq. \eqref{orc}:
\begin{equation}
    \gamma(x,y) =  1 - \frac{W_1(x,y)}{d_G(x,y)} = 1 - \frac{25/12}{1} = -\frac{13}{12} \ .
\end{equation}

\section{Block-encoding and quantum singular value transformation}
\label{sec: summaryofnecessarytechniques}
We briefly summarize the essential quantum tools used in our algorithm. For conciseness, we highlight only the main results and omit technical details, which are thoroughly covered in~\cite{gilyen2019quantum}. An identical summary is also presented in~\cite{lee2025new}.

\begin{definition}[Block-encoding unitary, see e.g.~\cite{low2017optimal, low2019hamiltonian, gilyen2019quantum}]
\label{def: blockencode} 
Let $A$ be a Hermitian matrix of size $N \times N$ with operator norm $\norm{A} < 1$. A unitary matrix $U$ is said to be an \emph{exact block encoding} of $A$ if
\begin{align}
    U = \begin{pmatrix}
       A & * \\
       * & * \\
    \end{pmatrix},
\end{align}
where the top-left block of $U$ corresponds to $A$. Equivalently, one can write
\begin{equation}
    U = \ket{\mathbf{0}}\bra{\mathbf{0}} \otimes A + (\cdots),    
\end{equation}
where $\ket{\mathbf{0}}$ denotes an ancillary state used for block encoding, and $(\cdots)$ represents the remaining components orthogonal to $\ket{\mathbf{0}}\bra{\mathbf{0}} \otimes A$. If instead $U$ satisfies
\begin{equation}
    U = \ket{\mathbf{0}}\bra{\mathbf{0}} \otimes \tilde{A} + (\cdots),
\end{equation}
for some $\tilde{A}$ such that $\|\tilde{A} - A\| \leq \epsilon$, then $U$ is called an {$\epsilon$-approximate block encoding} of $A$. Furthermore, the action of $U$ on a state $\ket{\mathbf{0}}\ket{\phi}$ is given by
\begin{align}
    \label{eqn: action}
    U \ket{\mathbf{0}}\ket{\phi} = \ket{\mathbf{0}} A\ket{\phi} + \ket{\mathrm{garbage}},
\end{align}
where $\ket{\mathrm{garbage}}$ is a state orthogonal to $\ket{\mathbf{0}}A\ket{\phi}$. The circuit complexity (e.g., depth) of $U$ is referred to as the {complexity of block encoding $A$}.
\end{definition}

Based on~\ref{def: blockencode}, several properties, though immediate, are of particular importance and are listed below.
\begin{remark}[Properties of block-encoding unitary]
The block-encoding framework has the following immediate consequences:
\begin{enumerate}[label=(\roman*)]
    \item Any unitary $U$ is trivially an exact block encoding of itself.
    \item If $U$ is a block encoding of $A$, then so is $\Ibb_m \otimes U$ for any $m \geq 1$.
    \item The identity matrix $\Ibb_m$ can be trivially block encoded, for example, by $\sigma_z \otimes \Ibb_m$.
\end{enumerate}
\end{remark}

Given a set of block-encoded operators, various arithmetic operations can be done with them. Here, we simply introduce some key operations that are especially relevant to our algorithm, focusing on how they are implemented and their time complexity, without going into proofs. For more detailed explanations, see~\cite{gilyen2019quantum, camps2020approximate}.

\begin{lemma}[Informal, product of block-encoded operators, see e.g.~\cite{gilyen2019quantum}]
\label{lemma: product}
    Given unitary block encodings of two matrices $A_1$ and $A_2$, with respective implementation complexities $T_1$ and $T_2$, there exists an efficient procedure for constructing a unitary block encoding of the product $A_1 A_2$ with complexity $T_1 + T_2$.
\end{lemma}

\begin{lemma}[Informal, tensor product of block-encoded operators, see e.g.~{\cite[Theorem 1]{camps2020approximate}}]\label{lemma: tensorproduct}
    Given unitary block-encodings $\{U_i\}_{i=1}^m$ of multiple operators $\{M_i\}_{i=1}^m$ (assumed to be exact), there exists a procedure that constructs a unitary block-encoding of $\bigotimes_{i=1}^m M_i$ using a single application of each $U_i$ and $\mathcal{O}(1)$ SWAP gates.
\end{lemma}

\begin{lemma}[Informal, linear combination of block-encoded operators, see e.g.~{\cite[Theorem 52]{gilyen2019quantum}}]
    Given the unitary block encoding of multiple operators $\{A_i\}_{i=1}^m$. Then, there is a procedure that produces a unitary block encoding operator of $\sum_{i=1}^m \pm (A_i/m) $ in time complexity $\mathcal{O}(m)$, e.g., using the block encoding of each operator $A_i$ a single time. 
    \label{lemma: sumencoding}
\end{lemma}

\begin{lemma}[Informal, Scaling multiplication of block-encoded operators] 
\label{lemma: scale}
    Given a block encoding of some matrix $A$, as in~\ref{def: blockencode}, the block encoding of $A/p$ where $p > 1$ can be prepared with an extra $\mathcal{O}(1)$ cost.
\end{lemma}



\begin{lemma}[Matrix inversion, see e.g.,~\cite{gilyen2019quantum, childs2017quantum}]\label{lemma: matrixinversion}
Given a block encoding of some matrix $A$  with operator norm $||A|| \leq 1$ and block-encoding complexity $T_A$, then there is a quantum circuit producing an $\epsilon$-approximated block encoding of ${A^{-1}}/{\kappa}$ where $\kappa$ is the conditional number of $A$. The complexity of this quantum circuit is $\mathcal{O}\left( \kappa T_A \log \left({1}/{\epsilon}\right)\right)$. 
\end{lemma}


\end{document}